\documentclass[aps,pre,twocolumn,groupedaddress,showpacs]{revtex4-2}
\usepackage{graphicx}
\usepackage{bm}
\usepackage{color}

\newcommand{\bec}[1]{\mbox{\boldmath $ #1$}}
\newcommand{\meanrho}{\overline{\rho}}
\newcommand{\meanT}{\overline{T}}
\newcommand{\meanE}{\overline{E}}

\begin{document}
\title{Compressibility effects in a turbulent transport of temperature field}
\author{I.~Rogachevskii}
\email{gary@bgu.ac.il}
\homepage{http://www.bgu.ac.il/~gary}
\author{N. Kleeorin}
\email{nat@bgu.ac.il}
\affiliation{
Department of Mechanical Engineering, Ben-Gurion University of the Negev, P. O. B. 653, Beer-Sheva
 8410530, Israel
 \\
Nordita, Stockholm University and KTH Royal Institute of Technology, 10691 Stockholm, Sweden}
\date{\today}
\begin{abstract}
Compressibility effects in a turbulent transport of temperature field
are investigated applying the quasi-linear approach
for small P\'eclet numbers and the spectral $\tau$ approach
for large P\'eclet numbers.
Compressibility of a fluid flow reduces the
turbulent diffusivity of the mean temperature field similarly to
that for particle number density and magnetic field.
However, expressions for the turbulent diffusion
coefficient for the mean temperature field in a compressible turbulence
are different from those for the mean particle number density
and the mean magnetic field.
Combined effect of compressibility and inhomogeneity of turbulence
causes an increase of the mean temperature
in the regions with more intense velocity fluctuations due to a turbulent pumping.
Formally, this effect is similar to a phenomenon of compressible turbophoresis
found previously [J. Plasma Phys. {\bf 84}, 735840502 (2018)]
for non-inertial particles or gaseous admixtures.
Gradient of the mean fluid pressure results in an additional
turbulent pumping of the mean temperature field.
The latter effect is similar to turbulent barodiffusion
of particles and gaseous admixtures.
Compressibility of a fluid flow also
causes a turbulent cooling of the surrounding fluid
due to an additional sink term in the equation for the mean temperature field.
There is no analog of this effect for particles.
\end{abstract}

\maketitle

\section{Introduction}

Compressibility of a fluid flow affects turbulent
transport of particles, temperature and magnetic fields
(see, e.g., \cite{MY71,MY75,ZRS90,Z08,IR21}),
e.g., it causes qualitative changes in the
properties of both, mean fields and fluctuations.
Large-scale effects of turbulence on particle concentrations and temperature field
are described by means of turbulent flux of particles
and turbulent heat flux, respectively.
For incompressible flow, main contribution to the turbulent fluxes
is determined by turbulent diffusion of particles and temperature field.
This corresponds to gradient turbulent transport of particles and temperature fields, e.g.,
the turbulent flux of particles is directed opposite
to gradient of the mean particle number density,
while the turbulent heat flux is directed opposite
to gradient of the mean fluid temperature.

Compressibility of a turbulent flow results in
a reduction of turbulent diffusivity of a mean particle number density
at small \cite{RB11} and large \cite{RKB18,IR21} P\'eclet numbers.
The P\'eclet number is the ratio of nonlinear to diffusion terms in the
equation for particle number density fluctuations.
Similar effect of the reduction of turbulent magnetic diffusivity
by compressible turbulence exists also
for the mean magnetic field at small \cite{KR80,RB11} and large \cite{RKB18,IR21}
magnetic Reynolds numbers.
The conclusion about the reduction of turbulent diffusivity
by the compressibility of fluid flow
has been also confirmed by the test-field method
in direct numerical simulations for an irrotational homogeneous
deterministic flow \cite{RB11}.
Various aspects related to compressibility effects
on turbulent transport have been studied using
different analytical approaches (see for a review, \cite{IR21}),
e.g., the quasi-linear approach \cite{KR80,RB11,RKB18},
the spectral tau approach \cite{RKB18}, the path-integral
approach \cite{EKR95,EKR96,EKR97}, the multiple-scale
direct-interaction approximation \cite{YO18,Y18}, etc.

Compressibility of a turbulent flow causes additional
non-gradient contribution to turbulent flux of particles that is
proportional to a product of the mean particle number density
and effective pumping velocity.
In a density stratified turbulence, the
effective pumping velocity of particles is proportional to
the gradient of the mean fluid density multiplied by turbulent diffusion coefficient
\cite{EKR96,EKR97}.
The pumping effect results in accumulation
of particles in regions of maximum mean fluid density.

In a temperature stratified turbulence,
similar effect referred as turbulent thermal diffusion
results in a turbulent non-diffusive
flux of particles in the direction of the turbulent heat flux, so that
particles are accumulated in the vicinity of the mean temperature minimum \cite{EKR96,EKR97}.
This phenomenon has been studied theoretically \citep{EKR00,EKR01,PM02,RE05,AEKR17},
found in direct numerical simulations \citep{HKRB12,BRK12,RKB18}, detected in
different laboratory experiments \citep{BEKR04,EEKR06,EEKR10,AEKR17},
and atmospheric turbulence with temperature inversions \citep{SEKR09}.
This effect has been shown to be important for concentrating dust
in protoplanetary discs \citep{H16}.
Density stratification which causes turbulent pumping of particles,
becomes weaker with increasing compressibility,
i.e., with increasing the Mach number \cite{RKB18}.

Compressibility of a fluid flow in inhomogeneous turbulence also results in a new pumping effect of particles
from regions of low to high turbulent intensity
both for small and large P\'eclet numbers.
This effect has been interpreted in \cite{RKB18} as a compressible turbophoresis
of non-inertial particles and gaseous admixtures, while the classical
turbophoresis effect for incompressible inhomogeneous turbulence
\cite{CTT75,R83,G97,EKR98,MHR18}
exists only for inertial particles and causes
them to be pumped to regions with lower turbulent intensity.

The compressibility of a turbulent fluid flow affects also passive scalar fluctuations.
In particular, it results in a slow scale-dependent turbulent diffusion of a small-scale
passive scalar fluctuations for large P\`{e}clet numbers \cite{EKR95}.
In addition, the level of the passive scalar fluctuations in the presence of
a gradient of the mean passive scalar field
in compressible turbulent flow can be fairly strong. On the other
hand, passive scalar transport in a density stratified turbulent fluid
flow is accompanied by formation of large-scale
structures due to instability of the
mean passive scalar field in inhomogeneous turbulent
velocity field \cite{EKR95}.

Another interesting feature for a compressible temperature stratified
turbulence is that turbulent flux of entropy is different from
turbulent convective flux of fluid internal energy
\cite{BR95,RK15}.
In particular,
in a low-Mach-number approximation as well as in the framework of the mean-field approach,
turbulent flux of entropy is given by ${\bm F}_s=\meanrho \,
\langle s' \, {\bm u} \rangle$, where $\meanrho$ is the mean fluid density and
$s'$ and ${\bm u}$ are fluctuations of entropy and velocity, respectively,
and the angular brackets $\langle ...\rangle$ denote ensemble averaging.
On the other hand, the turbulent convective flux of the fluid internal energy is
${\bm F}_c=\meanT \, \meanrho \, \langle s' \, {\bm u} \rangle$,
where $\meanT$ is the mean fluid temperature.
This turbulent convective flux is well-known in the astrophysical and geophysical literature,
and it cannot be used as a turbulent flux in the equation for the mean entropy.
This is exact result for low-Mach-number temperature stratified turbulence
and is independent of the turbulence model used \cite{RK15}.

Temperature fluctuations and anomalous scaling in a low-Mach-number
compressible turbulent flow have been studied in \cite{RKE97}.
Due to compressibility and external pressure fluctuations,
the anomalous scaling (i.e., the violation of the dimensional
analysis predictions for the scaling laws) may occur
in the second moment of the temperature field.
The cause of the anomalous behavior is a compressibility-induced
depletion of the turbulent diffusion of the
second moment of the temperature field \cite{RKE97}.

In spite of many studies of turbulent transport of passive scalar,
some large-scale (mean-field) features related to
compressibility effects on turbulent transport of temperature field are not known.
In the present paper, we study compressibility effects in turbulent transport
of the mean temperature field, i.e., we consider here mean-field effects.
This paper is organized as follows.
In Section II we outline the governing equations.
Turbulent heat flux and level of temperature fluctuations
are determined for small P\'eclet numbers in Section III
and for large P\'eclet numbers in Section IV.
In Sections III-IV we also outline the method of derivations
and approximations made for study of the compressibility effects.
In Section V we discuss how a homogeneous compressible turbulence
can cause a turbulent cooling of the surrounding fluid.
Finally, conclusions are drawn in Section VI.
In Appendix A we outline the multi-scale approach used in the present study.
Details of the derivation of turbulent heat flux and level of temperature fluctuations
are given in Appendix~B for small P\'eclet numbers
and in Appendix~C for large P\'eclet numbers.

\section{Governing equations}

Evolution of temperature field $T(t,{\bm r})$ in
a compressible fluid velocity field ${\bf U}(t,{\bm r})$ is given by \cite{LL87}
\begin{eqnarray}
{\partial T \over \partial t} + ({\bf U} \cdot {\bm  \nabla})
T + (\gamma - 1) T ({\bm \nabla} \cdot {\bm U})  = D \Delta T + J_\nu ,
\label{A1}
\end{eqnarray}
where $D$ is the molecular thermal conductivity, $\gamma=c_{\rm p}/c_{\rm v}$ is
the ratio of specific heats and $J_\nu$ is the heating source caused, e.g.,
by a viscous dissipation.

In a compressible flow, Eq.~(\ref{A1}) for the temperature field is different from
equation for particle number density $n(t,{\bm r})$ \cite{C43,AP81}
\begin{eqnarray}
{\partial n \over \partial t} + {\bm \nabla} \cdot (n \, {\bm U})  = D_n \, \Delta n ,
\label{A1b}
\end{eqnarray}
where $D_n$ is the coefficient of Brownian (molecular) diffusion of particles.

We consider a compressible turbulent flow when Mach number can be not small.
To derive equations for the turbulent heat flux
and the level of temperature fluctuations, we apply
the mean-field approach.
In particular, the fluid temperature, pressure, density
and velocity are decomposed into mean and fluctuating parts,
where the fluctuating parts have zero mean values,
i.e., the Reynolds averaging is applied here, which easily separates
fluctuations from mean fields. For example, the density-weighted
averaging quantities \cite{FA69,LE94} are usually difficult
to extract from laboratory and atmospheric measurements or
from astrophysical observations.

In the framework of the mean-field approach,
the fluid temperature is $T= \overline{T} + \theta$,
the fluid pressure is $P= \overline{P} + p$ and
the fluid density is $\rho= \overline{\rho} + \rho'$,
where $\overline{T}=\langle T \rangle$ is the mean fluid temperature,
$\overline{P}=\langle P \rangle$ is the mean fluid pressure
and $\overline{\rho}=\langle \rho \rangle$ is the mean fluid density,
$\theta$ are temperature fluctuations,
$p$ are pressure fluctuations and $\rho'$ are density fluctuations.
The angular brackets denote an ensemble averaging.
Similarly, ${\bm U}= \overline{\bm U} + {\bm u}$, where $\overline{\bm U}
=\langle {\bm U} \rangle$ is the mean fluid velocity, and ${\bm u}$ are
velocity fluctuations.
For simplicity, we consider the case $\overline{\bm U}=0$.

Averaging Eq.~(\ref{A1}) over ensemble of turbulent velocity field,
we arrive at equation for the mean temperature field as
\begin{eqnarray}
{\partial \overline{T} \over \partial t} + \bec{\nabla} {\bf \cdot}
\left\langle \theta  \, {\bm u} \right\rangle
= - (\gamma - 2) \, \left\langle \theta \, (\bec{\nabla} {\bf \cdot} {\bm u})\right\rangle
+ D \, \Delta \overline{T} + \overline{J}_\nu,
\nonumber\\
\label{A2}
\end{eqnarray}
where ${\bm F} = \left\langle \theta  \, {\bm u} \right\rangle$ is the turbulent heat flux,
$\overline{J}_\nu$ is the mean heating source caused
by the viscous dissipation of the turbulent kinetic energy, and $I_{\rm S} = - (\gamma - 2) \, \left\langle \theta \, (\bec{\nabla} {\bf \cdot} {\bm u})\right\rangle$ is the mean sink term resulting in a turbulent cooling
due to compressibility effects (see Section~V).
Using Eqs.~(\ref{A1}) and~(\ref{A2}), we obtain equation for
temperature fluctuations, $\theta({\bm x},t)=T-\overline{T}$:
\begin{eqnarray}
{\partial \theta \over \partial t} + {\cal Q} - D \bec\nabla \theta  = -  ({\bm u}{\bm \cdot} \bec{\nabla}) \overline{T} - (\gamma - 1) \,\overline{T} \, \bec\nabla {\bm \cdot} \, {\bm u} ,
\label{A3}
\end{eqnarray}
where
\begin{eqnarray*}
{\cal Q}=\bec\nabla {\bm \cdot} \, (\theta {\bm u}
- \langle {\bm u} \, \theta \rangle)
+ (\gamma - 2) \,[\theta \, \bec\nabla {\bm \cdot} \, {\bm u}
- \langle\theta \, \bec\nabla {\bm \cdot} \, {\bm u}\rangle]
\end{eqnarray*}
are nonlinear terms and
\begin{eqnarray*}
I = -  ({\bm u}{\bm \cdot} \bec{\nabla}) \overline{T}
- (\gamma - 1) \,\overline{T} \, \bec\nabla {\bm \cdot} \, {\bm u}
\end{eqnarray*}
are the source terms of temperature fluctuations.
The ratio of the nonlinear term to the diffusion
term is the P\'eclet number, that is estimated as
${\rm Pe} = u_{0} \, \ell_0 / D$,
where $u_{0}$ is the characteristic turbulent velocity in the integral
(energy-containing) scale $\ell_0$ of turbulence.
We consider a one way coupling, i.e., we take into account
the effect of turbulence on the temperature field, but neglect
the feedback effect of the temperature on the turbulence.

To determine the turbulent heat flux and the level of temperature fluctuations,
and to take into account small-scale properties of the turbulence,
we use two-point correlation functions.
For fully developed turbulence, scalings for
the turbulent correlation time and
the turbulent kinetic energy spectrum
are related via the Kolmogorov scalings \cite{MY71,MY75,Mc90,F95,P2000}.
We consider the cases with small and large P\'eclet and Reynolds numbers.

In the framework of the mean-field approach, we assume
that there is a separation of spatial and temporal scales, i.e.,
$\ell_0 \ll L_T$ and $\tau_0 \ll t_T$, where
$L_T$ and $t_T$ are the characteristic spatial and temporal scales
characterizing the variations of the mean temperature field and $\tau_0=\ell_0/u_0$.
The mean fields depend on ``slow'' variables, while fluctuations
depend on ``fast'' variables.
Separation into slow and fast variables is widely used in theoretical physics,
and all calculations are reduced to the Taylor expansions of all functions
using small parameters $\ell_0 / L_T$ and $\tau_0 / t_T$.
The findings are further truncated to leading order terms.
Separation to slow and fast variables
is performed by means of a standard multi-scale approach
\cite{RS75} discussed in details in Appendix~A.

\section{Turbulent heat flux and level of temperature fluctuations
for small P\'eclet numbers}

In this section we derive equations for the turbulent heat flux
and the level of temperature fluctuations for small P\'eclet numbers using the quasi-linear approach.
For a random flow with small P\'eclet and Reynolds numbers,
there are no universal scalings for the correlation time and
the turbulent kinetic energy spectrum.
This is the reason why we use non-instantaneous two-point correlation
functions in this case.
In the framework of the quasi-linear approach, we neglect the nonlinear term ${\cal Q}$, but keep
the molecular diffusion term in Eq.~(\ref{A3}).
We rewrite this equation in Fourier space
and find the solution of this equation, given by Eq.~(\ref{R1})
in Appendix~B.
Using this solution and applying the multi-scale approach (see \cite{RS75} and Appendix~A), we arrive at expressions for the turbulent heat flux and the level of temperature fluctuations
in Fourier space for small P\'{e}clet numbers as
\begin{widetext}
\begin{eqnarray}
\left\langle \theta  \, u_j \right\rangle &=& - {\gamma - 1 \over 2} \, \biggl[\overline{T} \, \int G_D \biggl(\nabla_i - 2 D k^2 G_D \, k_{im} \nabla_m + 2 i k_i \biggr) \, f_{ij} \,d {\bm k} \,d\omega
\nonumber\\
&& - \left(\nabla_i \overline{T}\right)  \, \int G_D \biggl(
{\gamma - 3 \over \gamma - 1}  \,\delta_{im} + 2 D k^2 G_D \, k_{im} + k_m {\partial \over \partial k_i} \biggr)f_{mj} \biggr] \,d {\bm k} \,d\omega   ,
\label{A6}\\
\left\langle  \theta^2 \right\rangle &=& {\gamma - 1 \over 4} \, \biggl\{\overline{T}
\, \int G_D \biggl[\biggl(2 D k^2 G_D \, k_{jn} \nabla_n -  \nabla_j \biggr)\, F_{j}^{(+)}
+ 2 i k_j \, F_{j}^{(-)}\biggr] \,d {\bm k} \,d\omega
\nonumber\\
&& + \left(\nabla_n \overline{T}\right)  \, \int G_D \biggl[{\gamma - 3 \over \gamma - 1}
\,\delta_{jn} + 2 D k^2 G_D \, k_{jn} + k_j {\partial \over \partial k_n}\biggr]
\, F_{j}^{(+)} \biggr\} \,d {\bm k} \,d\omega .
\label{AA6}
\end{eqnarray}
\end{widetext}
\noindent
Details of derivations of Eqs.~(\ref{A6}) and~(\ref{AA6}) are given in Appendix~B.
Here $G_D \equiv G_D({\bm k},\omega) = (D {\bm k}^2 + i \omega)^{-1}$,
$\, f_{ij}\equiv f_{ij}({\bm k},\omega) = \left\langle u_i({\bm k},\omega)
\, u_j(-{\bm k},-\omega)\right\rangle$,
$\, F_{j}^{(\pm)} = F_{j}({\bm k},\omega) \pm F_{j}(-{\bm k},\omega)$,
where $F_{j}({\bm k},\omega) = \left\langle \theta({\bm k},\omega)
\, u_j(-{\bm k},-\omega)\right\rangle$ is the turbulent heat flux
in Fourier space, $\delta_{ij}$ is the Kronecker unit tensor and $k_{ij}=k_{i} k_{j} /k^2$.
Since we consider a one way coupling, the correlation function
$f_{ij}$ in Eqs.~(\ref{A6}) and~(\ref{AA6}) should be replaced by $f_{ij}^{(0)}$
for the background random flow with zero turbulent heat flux.

We use a statistically stationary, density-stratified, inhomogeneous, compressible
and non-helical background random flow determined by the following correlation function
in Fourier space \cite{RKB18,IR21}:
\begin{widetext}
\begin{eqnarray}
f_{ij}^{(0)}({\bm k},\omega) &=&
{\Phi(\omega) \over 8 \pi \, k^2 \,
(1+ \sigma_c)} \biggl\{E(k) \, \biggl[\delta_{ij} - k_{ij}
+ {i \over k^2} \,\big(k_j \lambda_i - k_i \lambda_j\big) + {i \over 2 k^2} \,
\big(k_i \nabla_j - k_j \nabla_i\big) \biggr]
\nonumber\\
&& \quad + 2 \sigma_c \, E_c(k) \biggl[k_{ij} + {i \over 2 k^2}
\big(k_i \nabla_j - k_j \nabla_i\big) \biggr] \biggr\} \left\langle {\bm u}^2\right\rangle ,
\label{A7}
\end{eqnarray}
\end{widetext}
\noindent
where ${\bm \lambda} = - {\bm \nabla} \ln \meanrho$ characterizes the
fluid density stratification,
$\sqrt{\left\langle {\bm u}^2\right\rangle}$ is the characteristic turbulent
velocity at the maximum scale $\ell_{0}$ of random motions,
and the parameter
\begin{eqnarray}
\sigma_c = {\left\langle (\bec{\nabla} \cdot \, {\bm u})^2 \right\rangle
\over \left\langle(\bec{\nabla} \times {\bm u})^{2} \right\rangle}
  \label{A8}
\end{eqnarray}
is the degree of compressibility of the turbulent
velocity field.
We considered a weakly anisotropic background turbulence.
In particular, in derivation of Eq.~(\ref{A7}), we assumed that
$\ell_0 \ll H_\rho$ and $\ell_0 \ll L_u$,
where $L_u = \left|\bec{\nabla} \ln \left\langle {\bm u}^2\right\rangle\right|^{-1}$
is the characteristic scale of the inhomogeneity of turbulence, and
$H_\rho=|{\bm \lambda}|^{-1} = |{\bm \nabla} \ln \meanrho|^{-1}$ is the
mean density stratification scale, which is assumed to be constant.
These conditions allow us to take into account leading effects in Eq.~(\ref{A7}),
which are linear in stratification, $\propto |{\bm \lambda}|$, and inhomogeneity of turbulence,
$\propto |\bec{\nabla} \ln \left\langle {\bm u}^2\right\rangle|$.
We neglect in Eq.~(\ref{A7}) high-order effects which are of the order of
O$(\lambda^2 \, \left\langle {\bm u}^2\right\rangle)$, O$(\nabla^2 \, \left\langle {\bm u}^2\right\rangle)$,
O$(\lambda_i \nabla_i\left\langle {\bm u}^2\right\rangle)$.

Generally, stratification also contributes to div~${\bm u}$, i.e., it contributes to the parameter $\sigma_c$.
Since this contribution is small, i.e., it is of the order of $\sim$O$(\lambda^2 \, \left\langle {\bm u}^2\right\rangle)$, we neglect this contribution in Eq.~(\ref{A7}).
This allows us to separate effects of the arbitrary Mach number, characterized by
the parameter $\sigma_c$, and density stratification, described by ${\bm \lambda}$.
The degree of compressibility $\sigma_c$ depends on the Mach number.
This dependence is not known for arbitrary Mach numbers and can be determined, e.g., in direct numerical simulations.

In Eq.~(\ref{A7}), $E(k)$ and $E_c(k)$ are the
spectrum functions for incompressible and compressible parts of a random flow.
We assume that the random flow have a
power-law spectrum for incompressible $E(k) = (q-1) \, (k / k_{0})^{-q}\, k_0^{-1}$
and compressible $E_c(k) = (q_c-1) \, (k / k_{0})^{-q_c} \, k_0^{-1}$ parts,
where the wave number varies in the range, $k_0\leq k \leq k_\nu$. Here $k_{\nu} = 1 / \ell_{\nu}$
is the wave number based on the viscous scale $\ell_{\nu}$,
and $k_{0} = 1 / \ell_{0} \ll k_\nu$.
We assume also that there are no random motions for
$k < k_0$. In the model of a compressible background turbulence
used in \cite{RKB18}, the exponents $q=q_c$.
In the present study, we consider the case when the spectrum exponents of the incompressible and
compressible parts of random motions are different, i.e., $q\not =q_c$.

We assume that the frequency function $\Phi(\omega)$ has a
Lorentz profile, $\Phi(\omega)=[\pi \tau_0 \,
(\omega^2 + \tau_0^{-2})]^{-1}$, which corresponds to the
correlation function $\langle u_i(t) u_j(t+\tau)\rangle \propto \exp (-\tau / \tau_0)$.
Here the correlation time for small P\'eclet numbers $\tau_0 \equiv \ell_{0} / u_0
\gg (D k^2)^{-1}$ for all turbulent scales.
To derive Eq.~(\ref{A7}) we use identities given in Appendix~B.
Different contributions to Eq.~(\ref{A7}) have been discussed in
\cite{B53,EKR95,RKB18,IR21}.

Integration in $\omega$ and ${\bm k}$ space in Eq.~(\ref{A6}) yields an equation
for the turbulent heat flux for small P\'eclet numbers:
\begin{eqnarray}
\left\langle  \theta  \, {\bm u} \right\rangle = \overline{T} \, {\bm V}^{\rm eff} - D_{_{T}} \, \bec\nabla \overline{T} ,
\label{A9}
\end{eqnarray}
where the turbulent diffusivity $D_{_{T}}$ and the effective pumping
velocity ${\bm V}^{\rm eff}$ are given by
\begin{eqnarray}
D_{_{T}} &=&  {(q-1)\over 3\, (q+1)} \, {\tau_0 \, \left\langle {\bm u}^2\right\rangle \over 1 + \sigma_c} \, {\rm Pe} \left[\gamma - {1 \over 2} (3\gamma -5) \, \sigma_c \, C_\sigma  \right],
\nonumber\\
\label{A10}
\end{eqnarray}
\begin{eqnarray}
{\bm V}^{\rm eff} &=& (\gamma-1) \, {(q-1)\over 3(q+1)} \,  {\tau_0 \, \left\langle {\bm u}^2\right\rangle \over (1 + \sigma_c)}  \,  {\rm Pe} \biggl[{3 \over 2} \, C_\sigma \, \sigma_c \,  {\bm \lambda}_u
+ {\bm \lambda}_P \biggr] .
\nonumber\\
\label{A11}
\end{eqnarray}
Here ${\bm \lambda}_u = {\bm \nabla} \ln \left\langle {\bm u}^2\right\rangle$, ${\bm \lambda}_P = {\bm \nabla} \ln \overline{P}$, and
\begin{eqnarray}
C_\sigma &=& {(q_c-1) \, (q+1)\over (q_c+1) \, (q-1)} .
\label{A12}
\end{eqnarray}
We take into account that the equation of state for ideal gas yields
${\bm \lambda}=-{\bm \lambda}_P + {\bm \nabla} \ln \overline{T}$.
Since $\tau_0 \, {\rm Pe} = \ell_0^2 /D$, the turbulent transport coefficients
given by Eqs.~(\ref{A10}) and (\ref{A11}) are determined only by the microphysical
diffusion time scale $\ell_0^2 /D$ for small P\'eclet numbers.
Equation~(\ref{A10}) implies that for small
P\'eclet numbers, compressibility effects in most of the cases decrease
the turbulent diffusivity.
Indeed, for $\gamma \geq 5/3$, the derivative $\partial D_{_{T}} / \partial \sigma_c$
is always negative, i.e., the compressibility effects decrease
the turbulent diffusivity. When $1< \gamma < 5/3$, the derivative
$\partial D_{_{T}} / \partial \sigma_c$ is negative when $C_\sigma < 2 \gamma/(5 -3\gamma)$.
For example, for $q=q_c$ the derivative $\partial D_{_{T}} / \partial \sigma_c$
is always negative. When $q=5/3$ (i.e., for the Kolmogorov spectrum)
and $q_c=2$ (i.e., for the Burgers turbulence with shock waves),
the derivative $\partial D_{_{T}} / \partial \sigma_c$
is negative when $10/9 < \gamma < 5/3$.
Note that, the total diffusivity $D+D_{_{T}}$ cannot be negative,
because for ${\rm Pe} \ll 1$ the molecular diffusivity
is much larger than the turbulent one, $D \gg |D_{_{T}}|$.

The first term $(\propto \sigma_c \, {\bm \nabla} \left\langle{\bm u}^2\right\rangle)$ in Eq.~(\ref{A11}) for the effective pumping velocity ${\bm V}^{\rm eff}$
of the mean temperature field describes a combined effect of compressibility of fluid flow
and  inhomogeneity of turbulence. This effect increases the mean temperature field in the regions with more intense velocity fluctuations due to turbulent pumping.
This effect is similar to a phenomenon of compressible turbophoresis
found previously for non-inertial particles or gaseous admixtures \cite{RKB18}.

The second term $(\propto {\bm \lambda}_P)$ in Eq.~(\ref{A11}) describes an additional
turbulent pumping effect due to the gradient of the mean fluid pressure.
This effect is similar to turbulent barodiffusion
of particles and gaseous admixtures \cite{EKR97}.
The physics of these effects is discussed in the next section.
Note that the expressions for turbulent diffusion and the effective pumping velocity
for the mean temperature field in a compressible turbulence
are different from those for particle number density and magnetic field
(see \cite{RKB18}), because equations for particle number density or magnetic field
are different from those for the fluid temperature (see discussion at the end of Section~IV).

Integration in $\omega$ and ${\bm k}$ space in Eq.~(\ref{AA6}) yields the expression
for the level of temperature fluctuations for small P\'eclet numbers as
\begin{widetext}
\begin{eqnarray}
\left\langle  \theta^2 \right\rangle &=& (\gamma-1)^2 \, \left({q_c-1 \over q_c+1} \right)\, \left({\sigma_c \over 1 + \sigma_c} \right) \, {\rm Pe}^2 \, \, \overline{T}^2
\nonumber\\
&& + {q-1  \over 3 (q+3)} \, {\rm Pe}^2 \, \ell_0^2 \, \biggl\{\left({\bm \nabla}\overline{T}\right)^2
+ {1 \over 8} \, (\gamma-1) \, \Big[6 ({\bm \lambda} \cdot {\bm \nabla})
+ (\gamma+3) \,({\bm \lambda}_u \cdot {\bm \nabla}) \Big] \overline{T}^2 \biggr\} .
\label{A14}
\end{eqnarray}
\end{widetext}
\noindent
The first term in the right hand side of Eq.~(\ref{A14}) determines
a dominant contribution of the compressible part of velocity fluctuations
to the level of temperature fluctuations.
Here we neglect much smaller contributions $\sim {\rm O}[\ell_0^2/(L_T L_u)]$,
${\rm O}[\ell_0^2/(L_T H_\rho)]$,
${\rm O}[\ell_0^2/L_T^2]$, caused by the compressible part of velocity fluctuations,
where $L_T$ is the characteristic scale of the mean temperature field variations.
For small $\sigma_c$, the level of temperature fluctuations
is determined by the terms given by the second line of Eq.~(\ref{A14})
and caused by the mean temperature gradient and the density stratified
and inhomogeneous part of velocity fluctuations.

\section{Turbulent heat flux and level of temperature fluctuations
for large P\'eclet numbers}

In this section we determine the turbulent heat flux and the level of temperature fluctuations
for large P\'eclet and Reynolds numbers.
We consider fully developed turbulence, where the Strouhal number
is of the order of unity and the turbulent correlation time
is scale-dependent, so we apply the Fourier transformation only in
${\bm k}$ space.

The procedure of the derivations of the expressions for the turbulent heat flux and the level of temperature fluctuations includes: (i) derivation of equations for the second moments in
${\bm k}$ space using the multi-scale approach, (ii) application of the spectral $\tau$ approach (see below) which allows us to relate the deviations of the third moments (appearing due to nonlinear terms) from those of the background turbulence with the deviations of the second moments, (iii) solution of the equations for the second moments in the ${\bm k}$ space, and (iv) inverse transformation to the physical space to obtain formulas for the turbulent heat flux and the level of temperature fluctuations.

Starting with Eq.~(\ref{A3}) for the temperature fluctuations $\theta$ and the Navier-Stokes equation
for the velocity ${\bm u}$ written in Fourier space, we derive dynamic
equations for the turbulent heat flux and level of temperature fluctuations as
\begin{widetext}
\begin{eqnarray}
&& {\partial F_j \over \partial t} = - {1 \over 2} \, (\gamma-1) \, \biggl[\overline{T} \biggl(2i k_i + \nabla_i \biggr)f_{ij}
- \left(\nabla_i \overline{T}\right) \, \biggl({\gamma -3 \over \gamma-1} \, \delta_{im} + k_{m} \, {\partial  \over \partial k_{i}}\biggr) f_{mj} \biggr] + \hat{\cal M} F_j^{\rm(III)} ,
\label{C2}\\
&& {\partial E_{\theta} \over \partial t} = {1 \over 2} (\gamma-1)\,\biggl[\overline{T} \left(2 i k_j F_{j}^{(-)}
- \nabla_j F_{j}^{(+)} \right)
+ \left(\nabla_m \overline{T}\right) \, \left(k_j {\partial \over\partial k_m} + {\gamma -3 \over \gamma-1} \, \delta_{jm}\right) F_{j}^{(+)}\biggr]
+ \hat{\cal M} E_{\theta}^{\rm(III)} .
\label{D21}
\end{eqnarray}
\end{widetext}
\noindent
Details of derivations of Eqs.~(\ref{C2})--(\ref{D21}) are given in Appendix~C.
Here $F_{j}({\bm k}) = \left\langle \theta({\bm k}) \, u_j(-{\bm k})\right\rangle$,
$\, E_{\theta}({\bm k}) = \left\langle \theta({\bm k}) \, \theta(-{\bm k})\right\rangle$,
$\, f_{ij}({\bm k}) = \left\langle u_i({\bm k}) \, u_j(-{\bm k})\right\rangle$,
and $F_{j}^{(\pm)} = F_{j}({\bm k}) \pm F_{j}(-{\bm k})$,
the third-order moment terms $\hat{\cal M} F_j^{\rm(III)}$ and $\hat{\cal M} E_{\theta}^{\rm(III)}$
written in ${\bm k}$ space and appearing due to the nonlinear terms
are given by Eqs.~(\ref{C3}) and~(\ref{CC3}) in Appendix~C.

Equations~(\ref{C2}) and~(\ref{D21}) for the second moment
include first-order spatial differential
operators $\hat{\cal M}$ applied to the
third-order moments $F^{\rm(III)}$. The problem arises
how to close Eqs.~(\ref{C2}) and~(\ref{D21}), i.e., how to express
the third-order terms $\hat{\cal M} F^{\rm(III)}$
through the lower moments \citep{MY71,MY75,Mc90,O70}.
We use the spectral $\tau$ approach which
is a universal tool in turbulent transport
for strongly nonlinear systems.
The spectral $\tau$ approximation
postulates that the deviations of the third-moment terms, $\hat{\cal
M} F^{\rm(III)}({\bm k})$, from the contributions to
these terms afforded by the background
turbulence, $\hat{\cal M} F^{\rm(III,0)}({\bm k})$,
can be expressed through similar deviations
of the second moments, $F^{\rm(II)}({\bm k}) -
F^{\rm(II,0)}({\bm k})$ as
\begin{eqnarray}
&& \hat{\cal M} F^{\rm(III)}({\bm k}) - \hat{\cal M}
F^{\rm(III,0)}({\bm k})
= - {F^{\rm(II)}({\bm k}) - F^{\rm(II,0)}({\bm k}) \over \tau_r(k)} ,
\nonumber\\
\label{C4}
\end{eqnarray}
(see, e.g., \citep{O70,PFL76,KRR90}),
where $\tau_r(k)$ is the scale-dependent relaxation time which can be
identified with the correlation time $\tau(k)$ of the
turbulent velocity field for large fluid Reynolds
numbers and large P\'eclet numbers.
Here functions with superscript $(0)$ correspond to background
turbulence with zero turbulent heat flux.
Therefore, Eq.~(\ref{C4}) is reduced to
$\hat{\cal M} F_i^{\rm(III)}({\bm k}) = - F_i({\bm k})/ \tau(k)$
and $\hat{\cal M} E_\theta^{\rm(III)}({\bm k}) = - E_\theta({\bm k})/ \tau(k)$.
Validation of the $\tau$ approximation for different
situations has been performed in various numerical
simulations \citep{BS05,BK04,BSS05,BS05B,BRRK08,RKKB11,RB11,HKRB12,EKLR17,RKB18}.
We apply the $\tau$ approximation only to
study the deviations from the background turbulence which are
caused by the spatial derivatives of the mean temperature. The
background compressible inhomogeneous and density stratified turbulence
is assumed to be known (see below).

The $\tau$ approximation is a sort of the high-order closure
and in general is similar to Eddy Damped
Quasi Normal Markovian (EDQNM) approximation. However some principal
difference exists between these two approaches \cite{O70,PFL76}.
The EDQNM closures do not relax to equilibrium (the background turbulence),
and the EDQNM approach does not describe properly the motions in the
equilibrium state in contrast to the $\tau$ approximation.
Within the EDQNM theory, there is no dynamically determined
relaxation time, and no slightly perturbed steady state can be
approached. In the $\tau$ approximation, the
relaxation time for small departures from equilibrium is
determined by the random motions in the  equilibrium state, but
not by the departure from the equilibrium. As follows from
the analysis in \cite{O70}, the $\tau$ approximation describes
the relaxation to the equilibrium state (the background turbulence)
much more accurately than the EDQNM approach.

Next, we assume that the characteristic times of variation
of the second moments $F_i$ and $E_{\theta}$ are substantially
larger than the correlation time $\tau(k)$ in all turbulence scales.
This allows us to get steady-state solutions of Eqs.~(\ref{C2}) and~(\ref{D21}) as
\begin{widetext}
\begin{eqnarray}
&& \left\langle  \theta  \, u_j \right\rangle = - {1 \over 2} \, (\gamma-1) \, \int \tau(k) \, \biggl[\overline{T} \biggl(2i k_i + \nabla_i \biggr) f_{ij} - \left(\nabla_i \overline{T}\right) \, \biggl({\gamma -3 \over \gamma-1} \, \delta_{im} + k_{m} \, {\partial  \over \partial k_{i}}\biggr) f_{mj} \biggr] \, d{\bm k},
\label{C5}\\
&& \left\langle  \theta^2 \right\rangle = {1 \over 2} \, (\gamma-1)\, \int \tau(k) \, \biggl[ \overline{T} \left(2i k_j F_{j}^{(-)} - \nabla_j F_{j}^{(+)} \right)
+ \left(\nabla_m \overline{T}\right) \, \left(k_j {\partial \over\partial k_m} + {\gamma -3 \over \gamma-1} \, \delta_{jm}\right) F_{j}^{(+)} \biggr]  \, d{\bm k} .
\label{D22}
\end{eqnarray}
\end{widetext}
\noindent
In Eqs.~(\ref{C5}) and~(\ref{D22}) we take into account a one way coupling,
i.e., we neglect the effect of the mean temperature gradients
on the turbulent velocity field.
This implies that we replace the correlation function
$f_{ij}$ in Eqs.~(\ref{C5}) and~(\ref{D22}) by $f_{ij}^{(0)}$
for the background turbulent flow with zero turbulent heat flux.

We use statistically stationary, density-stratified, inhomogeneous,
compressible and non-helical background turbulence,
which is determined by the following correlation
function in ${\bm k}$ space \cite{RKB18,IR21}:
\begin{widetext}
\begin{eqnarray}
f_{ij}^{(0)}({\bm k}) &=&
{1 \over 8 \pi \, k^2 \,  (1+ \sigma_c)} \,
\biggl\{E(k) \, \biggl[\delta_{ij} - k_{ij}
+ {i \over k^2} \,\big(k_j \lambda_i - k_i \lambda_j\big) + {i \over 2 k^2} \,
\big(k_i \nabla_j - k_j \nabla_i\big) \biggr]
\nonumber\\
&&
+ 2 \sigma_c \, E_c(k) \biggl[k_{ij} + {i \over 2 k^2} \,
\big(k_i \nabla_j - k_j \nabla_i\big) \biggr] \biggr\} \left\langle {\bm u}^2\right\rangle .
\label{C6}
\end{eqnarray}
\end{widetext}
\noindent
We assume here that the background turbulence is of Kolmogorov type with
constant energy flux over the spectrum,
i.e., the velocity fluctuations spectrum for the incompressible part
of turbulence in the
range of wave numbers $k_0<k<k_\nu$ is
$E(k) = - d \bar \tau(k) / dk$, where the function $\bar \tau(k) =
(k / k_{0})^{1-q}$ with $1 < q < 3$ being the
exponent of the turbulent kinetic energy spectrum.
The condition $q>1$ corresponds to finite turbulent kinetic energy
for very large fluid Reynolds numbers, while $q<3$ corresponds to finite
dissipation of the turbulent kinetic energy at the viscous scale
\citep[see, e.g.,][]{MY71,MY75,Mc90,F95,P2000}.
Similarly, the turbulent kinetic energy spectrum
for the compressible part
of turbulence is $E_c(k) = - d \bar \tau_c(k) / dk$, where the function $\bar \tau_c(k) =
(k / k_{0})^{1-q_c}$ with $1 < q_c < 3$.
For instance, the exponent of the incompressible part of
the turbulent kinetic energy spectrum
$q=5/3$ (the Kolmogorov spectrum), while the exponent of the compressible part of
the spectrum $q_c=2$ (for the Burgers turbulence with shock waves).
The turbulent correlation time in ${\bm k}$ space is
\begin{eqnarray}
\tau(k) = {2\tau_0 \over 1+\sigma_c} \, \Big[\bar \tau(k) + \sigma_c \,\bar \tau_c(k)\Big] .
\label{C7}
\end{eqnarray}
Note that for fully developed Kolmogorov like turbulence,
$\sigma_c < 1$ \cite{CH13}.

Integration in ${\bm k}$-space in Eq.~(\ref{C5}) yields the turbulent heat flux $\left\langle  \theta  \, {\bm u} \right\rangle = \overline{T} \, {\bm V}^{\rm eff} - D_{_{T}} \, \bec\nabla \overline{T}$, where the turbulent diffusivity $D_{_{T}}$ and the effective pumping
velocity ${\bm V}^{\rm eff}$ of the mean temperature field
for large P\'eclet numbers are given by
\begin{widetext}
\begin{eqnarray}
D_T &=&  {\tau_0 \, \left\langle {\bm u}^2\right\rangle \over 3}  \, \biggl\{1 + {\gamma-1 \over 1 + \sigma_c}
\biggl[1 - {\sigma_c \,  \over 2(1 + \sigma_c)} \Big(\tilde C_{\sigma} \, q
+ \sigma_c \,(q_c-1)\Big) \biggr] \biggr\} ,
\label{C8}\\
{\bm V}^{\rm eff} &=& (\gamma-1) \, {\tau_0 \, \left\langle {\bm u}^2\right\rangle \over 3 \, (1 + \sigma_c)}
\, \biggl\{ {\sigma_c \over 2} \, \biggl[1 + {\tilde C_{\sigma} \over 2(1 + \sigma_c)} \biggr] \, {\bm \lambda}_u
+ \biggl[1 - {\tilde C_{\sigma} \, \sigma_c \over 2(1 + \sigma_c)} \biggr] \, {\bm \lambda}_P \biggr\} ,
\label{C9}
\end{eqnarray}
\end{widetext}
\noindent
and
\begin{eqnarray}
\tilde C_\sigma = {2 (q_c -1) \over q + q_c -2} .
\label{CC10}
\end{eqnarray}
Equation~(\ref{C8}) implies that for large
P\'eclet numbers, compressibility effects decrease
the turbulent diffusivity.
Indeed, the derivative $\partial D_{_{T}} / \partial \sigma_c$
is always negative when $\sigma_c(\tilde C_\sigma q - 2q_c) < \tilde C_\sigma q + 2$.
Since $\tilde C_\sigma > 0$ and $\tilde C_\sigma q - 2q_c < 0$
[the latter inequality is reduced to $(q_c-1)^2 + (q-1) > 0$],
the derivative $\partial D_{_{T}} / \partial \sigma_c$
is negative, i.e., compressibility effects do decrease
the turbulent diffusivity.

For irrotational flow ($\sigma_c \gg 1$), the turbulent diffusivity and the effective pumping
velocity for large P\'eclet numbers are given by
\begin{eqnarray}
D_T &=&  {1 \over 3} \tau_0 \, \left\langle {\bm u}^2\right\rangle \, \biggl[1 - {1 \over 2} (\gamma-1) \, (q_c -1) \biggr],
\label{RRTT20}\\
{\bm V}^{\rm eff} &=& \biggl({\gamma-1 \over 6}\biggr) \, \tau_0 \,  {\bm \nabla} \left\langle{\bm u}^2\right\rangle .
\label{RRTT10}
\end{eqnarray}
Equations~(\ref{C9}) and~(\ref{RRTT10}) determine effective pumping velocity ${\bm V}^{\rm eff}$
of the mean temperature field caused by the inhomogeneity of compressible turbulence
and the gradient of the fluid pressure.
Let us discuss mechanisms of the turbulent pumping effects.
The first term $(\propto \sigma_c \, {\bm \nabla} \left\langle{\bm u}^2\right\rangle)$ in Eq.~(\ref{C9}) implies that there is an additional contribution to the turbulent heat flux caused by the combined effect of the inhomogeneity of turbulence and compressibility of fluid flow.
This effect results in increase of the mean temperature in the region with more intense
velocity fluctuations in a compressible turbulence.
This effect can be understood
using the budget equation for the mean internal energy density
$\meanE= c_{\rm v} \meanT$, where $c_{\rm v}$ is the specific heat at constant volume.
In particular, one of the sources in the budget equation for the mean internal energy density
is $- \langle p {\bm \nabla} \cdot {\bm u}\rangle$ \cite{LL87},
so that $\partial (\meanrho \, \meanE) / \partial t \sim - \langle p {\bm \nabla} \cdot {\bm u}\rangle$,
where $p$ are pressure fluctuations.
As follows from the Bernoulli law,
variations of the sum $\delta (p + \rho {\bm u}^2/2) \approx 0$, so that $\delta p \approx - \delta(\rho {\bm u}^2/2)$. This implies that the mean internal energy (and the mean temperature) is larger in the region with more intense compressible velocity fluctuations.
The turbulent pumping effect of the mean temperature field
caused by the joint effect of compressibility and inhomogeneity of turbulence
is similar to a phenomenon of compressible turbophoresis
for non-inertial particles or gaseous admixtures \cite{RKB18}.
In particular, the expression for the effective pumping velocity for particles
due to the compressible turbophoresis is proportional to
${\bm V}^{\rm eff}_{\rm particles}  \propto \sigma_c \, \tau_0 \,  {\bm \nabla} \left\langle{\bm u}^2\right\rangle$.

The second term $(\propto  {\bm \nabla} \overline{P})$
in Eq.~(\ref{C9}) determines an additional contribution
to the turbulent heat flux caused by the gradient of the mean fluid pressure.
This turbulent pumping increases the mean temperature in the regions with
higher mean fluid temperature.
The mechanism of this effect is the following. Since there is an outflow of fluid from
the turbulent regions with higher mean fluid pressure,
the fluid density decreases in these regions and temperature increases.
This effect is similar to turbulent barodiffusion \cite{EKR97} of particles or gaseous admixtures.

Note that expressions~(\ref{A10}) and~(\ref{C8}) for turbulent diffusion coefficient of the
mean temperature field in a compressible turbulence are different from those for the mean particle
number density \cite{RKB18,IR21}. Indeed, Eq.~(\ref{A1}) for the temperature field contains an additional term, $(\gamma - 2) T$ div ${\bm U}$, in comparison with Eq.~(\ref{A1b}) for the particle number density.
Even for $\gamma = 2$ when this additional term vanishes and the equations for the
temperature field and the particle number density are similar, the expressions for turbulent
diffusion coefficient for the mean temperature field in a compressible turbulence are different
from those for the mean particle number density.

The main reason for this difference is as follows.
Particles in a fluid flow is a two-phase system,
while turbulent transport of fluid temperature is a one-phase system.
Equation~(\ref{A3}) for temperature fluctuations $\theta({\bm x},t)=T-\overline{T}$
has two source terms $I = -  ({\bm u}{\bm \cdot} \bec{\nabla}) \overline{T}
- (\gamma - 1) \,\overline{T} \, \bec\nabla {\bm \cdot} \, {\bm u}$,
where the first term $-({\bm u}{\bm \cdot} \bec{\nabla}) \overline{T}$ contributes
to turbulent diffusion $D_{_{T}}$, while the second term $- (\gamma - 1) \,\overline{T}
\, \bec\nabla {\bm \cdot} \, {\bm u}$ contributes to
the effective pumping velocity ${\bm V}^{\rm eff}$ of the mean temperature,
so that the turbulent heat flux in a compressible turbulence
is $\left\langle  \theta  \, {\bm u} \right\rangle = \overline{T} \, {\bm V}^{\rm eff}
- D_{_{T}} \, \bec\nabla \overline{T}$.
The contribution $\left({\bm V}^{\rm eff}\right)_{_{\bec\nabla \overline{T}}} =-D_{_{T}}^{\ast} \, {\bm \nabla} \overline{T} / \overline{T}$ to the effective pumping velocity ${\bm V}^{\rm eff}$ of the mean temperature
due to the mean temperature gradient $\bec\nabla \overline{T}$ is actually
an additional contribution to the turbulent diffusivity $D_{_{T}}$.
Indeed, we can rewrite this contribution as
\begin{eqnarray}
\overline{T} \, \left({\bm V}^{\rm eff}\right)_{_{\bec\nabla \overline{T}}} = \overline{T} \left(-D_{_{T}}^{\ast} \, {{\bm \nabla} \overline{T} \over \overline{T}}\right) = - D_{_{T}}^{\ast} {\bm \nabla} \overline{T} ,
\label{AA9}
\end{eqnarray}
where
\begin{eqnarray}
D_{_{T}}^{\ast} = (\gamma-1) \, {(q-1)\over 3(q+1)} \,  {\tau_0 \, \left\langle {\bm u}^2\right\rangle \over (1 + \sigma_c)}  \,  {\rm Pe} ,
\label{AA9a}
\end{eqnarray}
for ${\rm Pe} \ll 1$, and
\begin{eqnarray}
D_{_{T}}^{\ast} = (\gamma-1) \, {\tau_0 \, \left\langle {\bm u}^2\right\rangle \over 3 \, (1 + \sigma_c)} \biggl(1 - {\tilde C_{\sigma} \, \sigma_c \over 2(1 + \sigma_c)} \biggr) ,
\label{AA9b}
\end{eqnarray}
for ${\rm Pe} \gg 1$.
This is the main reason why the expressions for turbulent diffusion
coefficient for the mean temperature field in a compressible turbulence
are different from those for the mean particle number density.

Integration in ${\bm k}$-space in Eq.~(\ref{D22}) yields
the level of temperature fluctuations for large P\'eclet numbers
\begin{widetext}
\begin{eqnarray}
\left\langle  \theta^2 \right\rangle &=& 8 \, f_c \, (\gamma-1)^2  \left({\sigma_c \over 1 + \sigma_c} \right)^3 \overline{T}^2
+  {1 \over 9}  \, \ell_0^2 \, \biggl\{8\, \left({\bm \nabla}\overline{T}\right)^2
+ (\gamma-1) \, \Big[(\gamma+ 3) \, \left({\bm \lambda}_u \cdot {\bm \nabla}\right) + 2 \, (5 - \gamma) \, \left({\bm \lambda} \cdot {\bm \nabla}\right) \Big] \overline{T}^2\biggr\} ,
\label{DD1}
\end{eqnarray}
\end{widetext}
\noindent
where the function $f_c(q,q_c,\sigma_c)$ depends on the degree of compressibility and the exponents of spectra for the incompressible and compressible parts of velocity fluctuations:
\begin{eqnarray}
f_c &=& {q_c -1 \over 3 q_c -5} + {2 (q_c -1) \over \sigma_c (q + 2 q_c -5)} + {q_c -1 \over \sigma_c^2 (2 q + q_c -5)} .
\nonumber\\
\label{DD2}
\end{eqnarray}
The first term in the right hand side of Eq.~(\ref{DD1}) determines
a dominant contribution of the compressible part of velocity fluctuations
to the level of temperature fluctuations.
Here we neglect much smaller contributions $\sim {\rm O}[\ell_0^2/(L_T L_u)]$,
${\rm O}[\ell_0^2/(L_T H_\rho)]$,
${\rm O}[\ell_0^2/L_T^2]$, caused by the compressible part of velocity fluctuations.
For small $\sigma_c$, the level of temperature fluctuations
is determined by the other terms in Eq.~(\ref{DD1})
which are caused by the mean temperature gradient and the density stratified
and inhomogeneous part of velocity fluctuations.

\section{Turbulent cooling}

In this section we discuss how a homogeneous compressible turbulence can
cause a turbulent cooling of the surrounding fluid.
Equation~(\ref{A2}) for the mean temperature field $\overline{T}$ contains an additional sink term $I_{\rm S} = - (\gamma - 2) \, \left\langle \theta \, (\bec{\nabla} {\bf \cdot} {\bm u})\right\rangle$
which can result in the turbulent cooling of the surrounding fluid for $\gamma < 2$.
Indeed, substituting Eq.~(\ref{A9}) for the turbulent heat flux into Eq.~(\ref{A2}), we obtain the equation for the mean temperature field $\overline{T}$ as
\begin{eqnarray}
&& {\partial \overline{T} \over \partial t} + \bec\nabla {\bm \cdot} \,
\big[\overline{T} \, {\bm V}^{\rm eff} - (D+D_{_{T}}) \, \bec\nabla \overline{T} \big] = \overline{J}_\nu
\nonumber\\
&& \quad \quad - (\gamma - 2) \, \left\langle \theta \, (\bec{\nabla} {\bf \cdot}
{\bm u})\right\rangle ,
\label{C11}
\end{eqnarray}
where the sink term $I_{\rm S}$ in Eq.~(\ref{C11}) for small
P\'eclet numbers is given by
\begin{eqnarray}
I_{\rm S} = - (\gamma-1) \, (2 -\gamma) \, \left({\sigma_c \over 1 + \sigma_c} \right) \, {\rm Pe} \,  {\overline{T} \over \tau_0} ,
\label{AA50}
\end{eqnarray}
while for large P\'eclet numbers it is
\begin{eqnarray}
I_{\rm S} = - 6 \, (\gamma-1) \, (2 -\gamma) \, {\sigma_c \over (1 + \sigma_c)^2} \,  {\overline{T} \over \tau_0} \, \left[{\rm Re}^{1/4} + {\sigma_c \over 4} \ln {\rm Re} \right] .
\nonumber\\
\label{AA51}
\end{eqnarray}
In Eq.~(\ref{AA51}) for simplicity we determine $I_{\rm S}$ when that the exponent of the incompressible part of the turbulent kinetic energy spectrum  for large Reynolds numbers is
$q=5/3$, while the exponent of the compressible part of the spectrum is $q_c=2$
\cite{KN07,F13}.

Let us consider a simple case with a uniform mean temperature field.
The heating source $\overline{J}_\nu$ in Eq.~(\ref{C11}) caused
by the viscous dissipation of the turbulent kinetic energy is given by
\begin{eqnarray}
\overline{J}_\nu = {\nu \over c_{\rm v}} \left[\left\langle(\bec\nabla {\bm \times} \, {\bm u})^2\right\rangle + {4 \over 3} \left\langle(\bec\nabla {\bm \cdot} \, {\bm u})^2\right\rangle\right] ,
\label{AA52}
\end{eqnarray}
where $c_{\rm v}$ is the specific heat at constant volume.
Here we use the equation for the turbulent kinetic energy density
$E_K=\left\langle \rho \, {\bm u}^2\right\rangle/2$
for compressible turbulence written as
\begin{eqnarray}
{\partial E_K \over \partial t} + {\rm div} \, {\bm \Phi}_K
= - \varepsilon_{_{K}} + \Pi_{_{K}} ,
\label{M27}
\end{eqnarray}
where
\begin{eqnarray}
{\bm \Phi}_K &=& - \nu \, \left[\left\langle\rho \, {\bm u} {\bm \times} (\bec\nabla {\bm \times} \, {\bm u})\right\rangle + {4 \over 3} \left\langle \rho \, {\bm u} \,  (\bec\nabla {\bm \cdot} \, {\bm u}) \right\rangle\right]
\nonumber\\
&& + \left\langle{\bm u} \, \left(\rho \, {\bm u}^2 /2\right) \right\rangle
+ \left\langle {\bm u} \, p\right\rangle
\label{T9}
\end{eqnarray}
is the flux of the density of turbulent kinetic energy,
$p$ are fluid pressure fluctuations,
\begin{eqnarray}
\varepsilon_{_{K}} = \nu \, \left[\left\langle\rho \, (\bec\nabla {\bm \times} \, {\bm u})^2\right\rangle + {4 \over 3} \left\langle\rho \, (\bec\nabla {\bm \cdot} \, {\bm u})^2\right\rangle\right]
\label{T10}
\end{eqnarray}
is the dissipation rate of the density of turbulent kinetic energy, and $\Pi_{_{K}}=\langle\rho \, {\bm u} {\bm \cdot} {\bm f}\rangle + \left\langle p \,  (\bec\nabla {\bm \cdot} \, {\bm u}) \right\rangle$ is the production rate of the density of turbulent kinetic energy caused by the external force (e.g., by an external large-scale shear).
The production term includes also the pressure-dilatation term $\left\langle p \,  (\bec\nabla {\bm \cdot} \, {\bm u}) \right\rangle$ (see, e.g., \cite{AL13,PJ17}).
In the limit of low Mach numbers, the pressure-dilatation term in $\Pi_K$ is known to be much smaller than $\langle\rho \, {\bm u} {\bm \cdot} {\bm f}\rangle$, and hence it can be safely neglected.

Using Eq.~(\ref{C6}) for the second moment of velocity fluctuations in the background turbulence, we obtain that the viscous heating source $\overline{J}_\nu$ is given by
\begin{eqnarray}
\overline{J}_\nu = {\left\langle {\bm u}^2\right\rangle \over 2 \tau_0} (1 + \sigma_c)^{-1}
\left[1 + {8 \over 3} \sigma_c {\rm Re}^{-1/4}\right] .
\label{AA54}
\end{eqnarray}
Turbulence can generate acoustic waves, and
the rate of the energy radiated by the acoustic waves
per unit mass for small Mach numbers is \cite{L52,P52}
\begin{eqnarray}
E_{\rm w} = \alpha \, {\left\langle {\bm u}^2\right\rangle \over \tau_0} \, {\rm Ma}^5 ,
\label{M25}
\end{eqnarray}
where $\alpha \sim 10$--$10^2$ is numerical coefficient,
${\rm Ma}=u_{\rm rms}/c_{\rm s}$ is the Mach number, $u_{\rm rms}=\left\langle {\bm u}^2\right\rangle^{1/2}$ and
$c_{\rm s} = (\gamma \overline{P}/\overline{\rho})^{1/2}$
is the sound speed.
The second term in Eq.~(\ref{AA54}) describes compressibility contribution
to the rate of the viscous heating,
\begin{eqnarray}
\overline{J}_\nu^{\, ({\rm c})} = {4 \over 3} \, {\left\langle {\bm u}^2\right\rangle \over \tau_0}
\, {\sigma_c \over 1 + \sigma_c} \, {\rm Re}^{-1/4} .
\label{AA55}
\end{eqnarray}
Assuming that the compressibility contribution to
the viscous heating of turbulence $\overline{J}_\nu^{\,({\rm c})}$ is compensated by
the radiative wave energy density $E_{\rm w}$,
we obtain that the degree of compressibility for small Mach numbers is given by
\begin{eqnarray}
\sigma_c = {3\alpha \over 4} \,  {\rm Ma}^5 \, {\rm Re}^{1/4} .
\label{M26}
\end{eqnarray}
In the equilibrium, the total viscous heating $\overline{J}_\nu$ is compensated by the compressible cooling $I_{\rm S}$, so that the increase of the internal thermal energy caused by the viscous heating
is given by
\begin{eqnarray}
c_{\rm v} \, \overline{T}_c = {2\left\langle {\bm u}^2\right\rangle \over 9  \, \alpha \,  {\rm Ma}^5 \, {\rm Re}^{1/2}}  .
\label{AA56}
\end{eqnarray}
Taking into account that the sound speed $c_{\rm s}$
depends on the mean temperature, we obtain from Eq.~(\ref{AA56}) that the increase of the internal thermal energy caused by the viscous heating is given by
\begin{eqnarray}
c_{\rm v} \, \overline{T}_c = C_\ast \, \left\langle {\bm u}^2\right\rangle \, {\rm Re}^{1/3} ,
\label{AA57}
\end{eqnarray}
where $C_\ast=(9 \, \alpha/2)^{2/3} / [\gamma \,(\gamma-1)]^{5/3}$.
Equation~(\ref{AA57}) can be rewritten in terms of the Mach number ${\rm Ma}=u_{\rm rms}/c_{\rm s}$ as
\begin{eqnarray}
{\rm Ma} = \left[{2\gamma \,(\gamma-1) \over 9 \, \alpha} \right]^{1/3} \, {\rm Re}^{-1/6} .
\label{AA58}
\end{eqnarray}
For example, taking parameters typical for the atmospheric turbulence, $\ell_0 = 10^2$ cm and $u_{\rm rms} = 2.7 \times 10^2$ cm/s and $\alpha=10$,
we obtain that $T_c=286$ K.

\section{Discussion and conclusions}

In the present study we have investigated compressibility effects
on turbulent transport of the mean temperature field.
We use the quasi-linear approach for study turbulent transport for small P\'eclet numbers.
When nonlinear effects are much stronger than the molecular diffusion
(i.e., for large P\'eclet numbers), we apply the spectral $\tau$ approach.
Similarly to turbulent transport of particles and magnetic fields, the
compressibility decreases the turbulent diffusivity of the mean temperature field,
but the expression for turbulent diffusivity for the mean temperature field in a compressible turbulence
is different from those for turbulent diffusivity of the mean particle number density and
turbulent magnetic diffusivity of the mean magnetic field.

We have found also turbulent pumping of the mean temperature field
due to joint effects of the fluid flow compressibility
and inhomogeneity of turbulence.
This effect causes an increase of the mean temperature
in the regions of more intense velocity fluctuations.
Similar compressibility effect referred to compressible turbophoresis \cite{RKB18},
results in a pumping of non-inertial particles or gaseous admixtures
from regions of low to high turbulent intensity.
Turbulent pumping also can be due to the gradients of the mean fluid pressure
resulting in increase of the mean temperature in the regions with increased
mean fluid pressure, similarly to phenomenon of turbulent barodiffusion
of particles and gaseous admixtures.

Due to compressibility, there is an additional sink term in the equation
for the mean fluid temperature, causing a turbulent cooling
in homogeneous turbulence.
This implies that there can be an equilibrium in a compressible homogeneous
turbulence with a uniform mean fluid temperature,
where the heating caused by the viscous dissipation of the turbulent kinetic energy
can be compensated by the turbulent cooling caused by the fluid compressibility.
Such effect does not exist in turbulent transport of particles or gaseous admixtures
in a compressible fluid flow.

To derive expressions for the turbulent heat flux and
the level of temperature fluctuations for large
P\'eclet and Reynolds numbers
in a compressible inhomogeneous and density stratified turbulence,
we apply the spectral $\tau$ approach (see Sect.~IV).
The $\tau$ approach reproduces many well-known phenomena found
by other methods in turbulent transport of particles, temperature and
magnetic fields, in turbulent convection and stably stratified
turbulent flows (see for a review, \cite{IR21}).
In turbulent transport, the $\tau$ approach yields correct
formulae for turbulent diffusion, turbulent thermal diffusion
and turbulent barodiffusion \cite{EKR95,EKR96,EKR97,BF03}.
The phenomenon of turbulent thermal diffusion
was predicted using the stochastic calculus (the path
integral approach). This effect was also reproduced
using the quasi-linear approach, the spectral $\tau$ approach
and the renormalization approach.

The $\tau$ approach reproduces the
well-known $k^{-7/3}$ spectrum of anisotropic velocity fluctuations
in a sheared turbulence (see \cite{EKRZ02}).
This spectrum was previously found in analytical, numerical,
laboratory studies and was observed in the atmospheric
turbulence (see, e.g., \cite{L67}).
In the turbulent boundary layer problems, the
$\tau$ approach yields correct expressions for turbulent viscosity,
turbulent thermal conductivity and the turbulent heat flux \cite{MY75,Mc90}.
This approach also describes the counter wind turbulent heat flux
and the Deardorff's heat flux in convective boundary layers (see \cite{EKRZ02}).
These phenomena were studied using different approaches
(see, e.g.,  \cite{MY75,Mc90,Z91}).

In magnetohydrodynamics, the $\tau$ approach reproduces many
well-known phenomena found by different methods, e.g., the
$\tau$ approximation yields correct formulae
for the $\alpha$-effect, the turbulent
diamagnetic and paramagnetic velocities,
the turbulent magnetic diffusion,
the ${\bf \Omega} {\bf \times} {\bf J}$ effect and
the $\kappa$-effect \cite{KR80,RK00,RKR03}.

\medskip

\begin{acknowledgements}
We have benefited from stimulating discussions with Axel Brandenburg and Michael Liberman.
This research was supported in part by Ministry of Science and Technology (grant No. 3-16516)
and PAZY Foundation of the Israel Atomic Energy Commission (IAEC) (grant No. 122-2020).
\end{acknowledgements}

\appendix

\section{Multi-scale approach}

In the framework of the multi-scale approach \cite{RS75},
the non-instantaneous two-point second-order correlation
functions are written as follows:
\begin{widetext}
\begin{eqnarray}
\left\langle \theta({\bm x},t_1) \, u_j ({\bm  y},t_2)\right\rangle
&=& \int \left\langle \theta({\bm k}_1,\omega_1) u_j({\bm k}_2,\omega_2)\right\rangle \, \exp
\big[i({\bm  k}_1 {\bm \cdot} {\bm x}
+{\bm k}_2 {\bm \cdot}{\bm y}) + i(\omega_1 t_1 + \omega_2 t_2)\big] \,d\omega_1 \, d\omega_2 \,d{\bm k}_1 \, d{\bm k}_2
\nonumber\\
&=& \int F_{j}({\bm k},\omega,t,{\bm R})  \exp[i {\bm k}
{\bm \cdot} {\bm r} + i\omega \, \tilde \tau] \,d\omega \,d
{\bm k} ,
\label{P1}\\
\left\langle \theta({\bm x},t_1) \, \theta({\bm  y},t_2)\right\rangle
&=& \int \left\langle \theta({\bm k}_1,\omega_1) \theta({\bm k}_2,\omega_2)\right\rangle \, \exp
\big[i({\bm  k}_1 {\bm \cdot} {\bm x}
+{\bm k}_2 {\bm \cdot}{\bm y}) + i(\omega_1 t_1 + \omega_2 t_2)\big] \,d\omega_1 \, d\omega_2 \,d{\bm k}_1 \, d{\bm k}_2
\nonumber\\
&=& \int E_\theta({\bm k},\omega,t,{\bm R})  \exp[i {\bm k}
{\bm \cdot} {\bm r} + i\omega \, \tilde \tau] \,d\omega \,d{\bm k} ,
\label{PP1}
\end{eqnarray}
\end{widetext}
\noindent
where
\begin{eqnarray}
&& F_{j}({\bm k},\omega,{\bm R},t) =
 \int \left\langle \theta({\bm k}_1,\omega_1) \, u_j({\bm k}_2,\omega_2)\right\rangle \, \exp[i \Omega t
\nonumber\\
&& \quad + i {\bm K} {\bm \cdot} {\bm R}] \,d \Omega \,d {\bm  K}.
\label{A5}
\end{eqnarray}
\begin{eqnarray}
&& E_\theta({\bm k},\omega,{\bm R},t) =
 \int \left\langle \theta({\bm k}_1,\omega_1) \, \theta({\bm k}_2,\omega_2)\right\rangle \, \exp[i \Omega t
\nonumber\\
&& \quad + i {\bm K} {\bm \cdot} {\bm R}] \,d \Omega \,d {\bm  K}.
\label{AA5}
\end{eqnarray}
Here we introduce large-scale variables: ${\bm R} = ({\bm x}
+ {\bm y}) / 2$, $\, {\bm K} = {\bm k}_1 + {\bm k}_2$,
$\, t = (t_1 + t_2) / 2$, $\, \Omega = \omega_1 + \omega_2$,
and small-scale variables: ${\bm r} = {\bm x} - {\bm y}$,
$\, {\bm k} = ({\bm k}_1 - {\bm k}_2) / 2$, $\, \tilde \tau
= t_1 - t_2$, $\, \omega = (\omega_1 - \omega_2) / 2$.
This implies that $\omega_1 = \omega + \Omega / 2$, $\, \omega_2 =
- \omega + \Omega / 2$, ${\bm k}_1 = {\bm k} + {\bm  K} / 2$,
and ${\bm k}_2 = - {\bm k} + {\bm  K} / 2$.
Mean-fields depend on the large-scale variables, while
fluctuations depend on the small-scale variables.
Similarly to Eqs.~(\ref{P1})--(\ref{AA5}), correlation function for velocity
fluctuations reads
\begin{eqnarray}
&& f_{ij}({\bm k},\omega,{\bm R},t) =
 \int \left\langle u_i({\bm k}_1,\omega_1) \, u_j({\bm k}_2,\omega_2)\right\rangle
\, \exp[i \Omega t
\nonumber\\
&& \quad + i {\bm K} {\bm \cdot} {\bm R}] \,d \Omega \,d {\bm  K} .
\label{P3}
\end{eqnarray}
After separation into slow and fast variables and calculating the functions
$F_{j}({\bm k},\omega,{\bm R},t)$ and $E_\theta({\bm k},\omega,{\bm R},t)$,
Eqs.~(\ref{P1}) and~(\ref{PP1}) in the limit of ${\bm r} \to {\bm 0}$
and $\tilde \tau \to 0$ allow us to determine the turbulent flux of the temperature field
and the level of temperature fluctuations in physical space:
\begin{eqnarray}
\left\langle \theta({\bm x},t) \, u_j ({\bm  x},t)\right\rangle = \int F_{j}({\bm k},\omega,{\bm R},t)
\,d\omega \,d {\bm k} ,
\label{A4}
\end{eqnarray}
\begin{eqnarray}
\left\langle \theta({\bm x},t) \, \theta({\bm  x},t)\right\rangle = \int E_\theta({\bm k},\omega,{\bm R},t)
\,d\omega \,d {\bm k} .
\label{AA4}
\end{eqnarray}

\section{Derivation of Eqs.~(\ref{A6})--(\ref{A7})}

We rewrite Eq.~(\ref{A3}) in Fourier space
and find solution of this equation as
\begin{eqnarray}
&& \theta({\bm k}, \omega) = - i \biggl[(\gamma-1)\,
\int \overline{T}({\bm Q}) \, (k_i - Q_i)  \, u_i({\bm k} - {\bm Q}, \omega) \,d{\bm Q}
\nonumber\\
&& \quad + \int Q_i \, \overline{T}({\bm Q}) \, u_i({\bm k} - {\bm Q}, \omega) \,d{\bm Q}
\biggr] \,  G_D({\bm k}, \omega) ,
 \label{R1}
\end{eqnarray}
where $G_D({\bm k},\omega) = (D {\bm k}^2 + i \omega)^{-1}$.
Using Eqs.~(\ref{A4}) and~(\ref{R1}), we determine the functions $F_{j}({\bm k},{\bm R})$
and $E_\theta({\bm k},{\bm R})$ as
\begin{widetext}
\begin{eqnarray}
F_{j}({\bm k},{\bm R}) &=&  - i \int \left[(\gamma-1)\,\left(k_i + {K_{i} \over 2} - Q_i\right) + Q_i \right]
\, G_D({\bm k} + {\bm  K} / 2) \,  \left\langle u_i ({\bm k} + {\bm  K} / 2 - {\bm  Q}) u_j(-{\bm k} + {\bm  K}  / 2)\right\rangle
\nonumber\\
&& \times  \, \overline{T}({\bm Q}) \, \exp(i {\bm K} {\bm \cdot} {\bm R})\,d {\bm  K} \,d {\bm  Q} ,
\label{R2}\\
E_\theta({\bm k},{\bm R}) &=& - {i \over 2} \int \biggl\{
\biggl[(\gamma-1)\,\biggl(k_i + {K_{i} \over 2} - Q_i\biggr) + Q_i \biggr]
\, G_D({\bm k} + {\bm  K} / 2) \,  \left\langle \theta(-{\bm k} + {\bm  K}  / 2) \, u_i ({\bm k} + {\bm  K} / 2 - {\bm  Q}) \right\rangle
\nonumber\\
&& + \biggl[(\gamma-1)\,\biggl(-k_i + {K_{i} \over 2} - Q_i\biggr) + Q_i \biggr]\, G_D(-{\bm k} + {\bm  K} / 2) \,  \left\langle \theta({\bm k} + {\bm  K}  / 2) \, u_i (-{\bm k} + {\bm  K} / 2 - {\bm  Q}) \right\rangle \biggr\}
\nonumber\\
&& \times \overline{T}({\bm Q}) \, \exp(i {\bm K} {\bm \cdot} {\bm R}) \,d {\bm  K} \,d {\bm  Q} ,
\label{RR2}
\end{eqnarray}
\end{widetext}
\noindent
where the functions $F_{j}$, $G_D$ and $u_i$ depend also on $\omega$, and
$\overline{T}$ depend on $t$ as well.
To simplify the notations, we do not show these dependencies here.
To determine $f_{ij}({\bm k},{\bm  K},{\bm  Q}) =\left\langle u_i ({\bm k} + {\bm  K} / 2 - {\bm  Q}) u_j(-{\bm k} + {\bm  K}  / 2)\right\rangle$, we use the following new variables:
\begin{eqnarray}
\tilde {\bm k} &=& (\tilde {\bm k}_{1} - \tilde {\bm k}_{2}) / 2 =
{\bm k} - {\bm  Q} / 2 ,
\label{R4}
\end{eqnarray}
\begin{eqnarray}
\tilde {\bm K} &=& \tilde {\bm k}_{1}
+ \tilde {\bm k}_{2} = {\bm  K} - {\bm  Q} ,
\label{R5}
\end{eqnarray}
where
\begin{eqnarray}
\tilde {\bm k}_{1} &=& {\bm k} + {\bm  K} / 2 - {\bm  Q} \;, \quad
\tilde {\bm k}_{2} = - {\bm k} + {\bm  K} / 2 .
\label{R3}
\end{eqnarray}
Since $|{\bm Q}| \ll |{\bm k}|$ and $|{\bm K}| \ll |{\bm k}|$, we use the Taylor expansion
\begin{eqnarray}
&& f_{ij}({\bm k} - {\bm Q}/2, {\bm  K} - {\bm  Q}) =
f_{ij}({\bm k},{\bm  K} - {\bm  Q}) - \frac{1}{2}
{\partial f_{ij}\over\partial k_m} Q_m
\nonumber\\
&& \quad + O({\bm Q}^2) ,
\label{R6}
\end{eqnarray}
\begin{eqnarray}
&& G_D({\bm k} + {\bm  K} / 2) = G_D({\bm k}) \left[1 - D ({\bm k} \cdot {\bm  K}) G_D({\bm k}) \right]
\nonumber\\
&& \quad + O({\bm K}^2) .
\label{R7}
\end{eqnarray}
In the similar way we calculate other terms in Eqs.~(\ref{R2})--(\ref{RR2}).
Using Eqs.~(\ref{R2})--(\ref{R7}), we arrive at expressions~(\ref{A6})--(\ref{AA6}) for the
turbulent heat flux and the level of temperature fluctuations
in Fourier space for small P\'{e}clet numbers.

To derive Eq.~(\ref{A7}), the second rank tensor $f_{ij}^{(0)}$ is constructed as a linear combination of symmetric tensors, $\delta_{ij}$ and $k_{ij}$, with respect to the indexes $i$ and $j$,
and non-symmetric tensors: $k_i \lambda_j$, $ k_j \lambda_i$, $k_i \nabla_j
\left\langle {\bm u}^2\right\rangle$ and $ k_j \nabla_i
\left\langle {\bm u}^2\right\rangle$.
We consider here only linear effects in ${\bm \lambda}$ and ${\bm \nabla} \left\langle {\bm u}^2\right\rangle$.
To determine unknown coefficients multiplying by these tensors,
we use the following conditions in the derivation of Eq.~(\ref{A7}):
$\left\langle {\bm u}^2\right\rangle = \int f_{ii}^{(0)} ({\bm k},\omega,{\bm K}) \, \exp(i {\bm K} {\bm \cdot} {\bm R}) \, d {\bm k} \, d\omega \, d {\bm K}$, $f_{ij}^{(0)} ({\bm k},\omega,{\bm K}) = f_{ji}^{*(0)} ({\bm k},\omega,{\bm K}) = f_{ji}^{(0)} (-{\bm k},\omega,{\bm K})$, and
\begin{widetext}
\begin{eqnarray}
&&\left\langle \left({\rm div} \, {\bm u}\right)^2 \right\rangle = \int (k_i + K_i/2) \, (k_j - K_j/2) \, f_{ij}^{(0)} ({\bm k},\omega,{\bm K}) \, \exp(i {\bm K} {\bm \cdot} {\bm R})
\, d {\bm k} \, d\omega \, d {\bm K} .
\label{L10}
\end{eqnarray}
\end{widetext}
\noindent
The normalization conditions for the functions $\Phi(\omega)$, $E(k)$ and $E_c(k)$ in Eq.~(\ref{A7}) are $\int_{-\infty}^{\infty} \Phi(\omega) \,d\omega=1$, $\int_{k_0}^{k_d} E(k) \,dk=1$
and $\int_{k_0}^{k_d} E_c(k) \,dk=1$.
For very low Mach numbers, i.e., when the parameter $\sigma_c$ is very small,
the continuity equation can be written in the anelastic approximation,
${\rm div} \, (\meanrho \, {\bm u}) = 0$,
which implies that
$(ik_i + iK_i/2 - \lambda_i) f_{ij}^{(0)}({\bm k},\omega,{\bm K}) = 0$ and
$(-ik_j + iK_j/2 - \lambda_j) f_{ij}^{(0)} ({\bm k},\omega,{\bm K})= 0$.

For the integration over $\omega$ in Eqs.~(\ref{A6}) and~(\ref{AA6}), we use the following identities:
\begin{eqnarray*}
&& \int_{-\infty}^{\infty} {d\omega \over (\pm i \omega + D k^2)
\, (\omega^2 + \tau_0^{-2})} ={\pi \, \tau_0 \over \tau_0^{-1} + D \, k^2}
\approx {\pi \, \tau_0 \over D \, k^2} ,
\\
&& \int_{-\infty}^{\infty} {d\omega \over (i \omega + D k^2)
\, (- i \omega + D k^2) \, (\omega^2 +\tau_0^{-2})}
\nonumber\\
&& \quad ={\pi \, \tau_0 \over D \, k^2 \left(\tau_0^{-1} + D \, k^2\right)}
\approx {\pi \, \tau_0 \over (D \, k^2)^2} ,
\end{eqnarray*}
which are determined in the limit when
the correlation time $\tau_0 \gg (D^{(\theta)}
k^2)^{-1}$.
For the integration over angles in ${\bm k}$ space in Eqs.~(\ref{A6}) and~(\ref{AA6}),
we use the following identity:
\begin{eqnarray*}
&&\int_{0}^{2\pi} \, d\varphi \int_{0}^{\pi} \sin \vartheta \,d\vartheta \,
{k_i \, k_j \over k^2} = {4 \pi \over 3} \, \delta_{ij} .
\end{eqnarray*}
For the integration over $k$ in Eqs.~(\ref{A6}) and~(\ref{AA6}),
we use the following identities:
\begin{eqnarray*}
&&\int_{k_0}^{k_d} {E(k) \over k^2} \,dk = {q-1 \over q+1}\,  \ell_0^2,
\\
&&\int_{k_0}^{k_d} {E(k) \over k^4} \,dk = {q-1 \over q+3}\, \ell_0^4 .
\end{eqnarray*}

\section{Derivation of Eqs.~(\ref{C2}) and~(\ref{D21})}

In this Appendix we derive Eqs.~(\ref{C2}) and~(\ref{D21})
for large P\'eclet and Reynolds numbers.
Using Eq.~(\ref{A3}) for the temperature fluctuations $\theta$ and the Navier-Stokes equation
for the velocity ${\bm u}$ written in Fourier space, we derive
equations for the following correlation functions:
\begin{eqnarray}
F_{j}({\bm k},{\bm R}) &=&\int \left\langle\theta({\bm k}
+ {\bm  K} / 2) \,  u_j(-{\bm k} + {\bm  K} / 2)\right\rangle
\nonumber\\
&&\times \exp[i {\bm K} {\bm \cdot} {\bm R}] \,d {\bm  K} ,
\label{C1}
\end{eqnarray}
\begin{eqnarray}
E_{\theta}({\bm k},{\bm R}) &=&\int \left\langle\theta({\bm k}
+ {\bm  K} / 2) \,  \theta(-{\bm k} + {\bm  K} / 2)\right\rangle
\nonumber\\
&&\times \exp[i {\bm K} {\bm \cdot} {\bm R}] \,d {\bm  K} .
\label{D10}
\end{eqnarray}
For brevity of notations we omit the large-scale variable $t$
in the functions $F_{j}({\bm k},{\bm R},t)$, $E_{\theta}({\bm k},{\bm R},t)$
and the mean temperature $\overline{T}({\bm R},t)$.

To derive evolution equations in the Fourier space for
the turbulent heat flux $F_{j}({\bm k},{\bm R})$
and the level of temperature fluctuations $E_{\theta}({\bm k},{\bm R})$,
we rewrite Eq.~(\ref{A3}) for the temperature fluctuations in ${\bm k}$ space as
\begin{eqnarray}
&& {\partial \theta({\bm k}) \over \partial t} = - i \biggl[(\gamma-1)\,
\int \overline{T}({\bm Q}) \, (k_i - Q_i)  \, u_i({\bm k} - {\bm Q}) \,d{\bm Q}
\nonumber\\
&& \quad + \int Q_i \, \overline{T}({\bm Q}) \, u_i({\bm k} - {\bm Q}) \,d{\bm Q}
\biggr] - {\cal Q}({\bm k}) ,
\label{R10}
\end{eqnarray}
where ${\cal Q}({\bm k})$ are the nonlinear terms written in ${\bm k}$ space.
For brevity of notations we omit below the variable $t$ in the functions
$\overline{T}({\bm Q},t)$, $\theta({\bm k},t)$, $\theta^{\rm(N)}({\bm k},t)$ and $u_i({\bm k},t)$.

Using Eq.~(\ref{R10}) for the temperature fluctuations $\theta$
written in Fourier space, we derive
equations for the instantaneous two-point correlation functions $F_{j}({\bm k},{\bm R})$
and $E_{\theta}({\bm k},{\bm R})$ defined by Eqs.~(\ref{C1}) and~(\ref{D10}).
To this end we use the identities:
\begin{eqnarray}
&& {\partial \over \partial t} \left\langle \theta({\bm k}_1,t) \,
u_j({\bm k}_2,t)\right\rangle = \left\langle {\partial \theta({\bm k}_1,t) \over \partial t} \,
u_j({\bm k}_2,t)\right\rangle
\nonumber\\
&&  \quad + \left\langle \theta({\bm k}_1,t) \, {\partial u_j({\bm k}_2,t) \over \partial t}\right\rangle ,
\label{R11}
\end{eqnarray}
\begin{eqnarray}
&& {\partial \over \partial t} \left\langle \theta({\bm k}_1,t) \,
\theta({\bm k}_2,t)\right\rangle = \left\langle {\partial \theta({\bm k}_1,t) \over \partial t} \,
\theta({\bm k}_2,t)\right\rangle
\nonumber\\
&&  \quad + \left\langle \theta({\bm k}_1,t) \, {\partial \theta({\bm k}_2,t) \over \partial t}\right\rangle .
\label{D11}
\end{eqnarray}
Equations~(\ref{R10})--(\ref{D11}) yield the dynamic equations as
\begin{eqnarray}
{\partial F_{j}({\bm k},{\bm R}) \over \partial t} &=& J_{j}({\bm k},{\bm R}) + \hat{\cal M} F_j^{\rm(III)}({\bm k},{\bm R}) ,
\label{R12}
\end{eqnarray}
\begin{eqnarray}
{\partial E_{\theta}({\bm k},{\bm R}) \over \partial t} &=& S({\bm k},{\bm R}) + \hat{\cal M} E_{\theta}^{\rm(III)}({\bm k},{\bm R}) ,
\label{D12}
\end{eqnarray}
where
\begin{eqnarray}
&& \hat{\cal M} F_j^{\rm(III)}({\bm k},{\bm R}) = \int \biggl[\, \left\langle\theta({\bm k}_1) \,
{\partial u_j({\bm k}_2) \over \partial t}\right\rangle
\nonumber\\
&& \quad
- \left\langle {\cal Q}({\bm k}_1) u_j({\bm k}_2)\right\rangle\biggr]
\exp[i {\bm K} {\bm \cdot} {\bm R}] \,d {\bm  K} ,
\label{C3}\\
&& \hat{\cal M} E_{\theta}^{\rm(III)}({\bm k},{\bm R}) = - \int \biggl[\, \left\langle\theta({\bm k}_1) \, {\cal Q}({\bm k}_2)\right\rangle
\nonumber\\
&& \quad - \left\langle {\cal Q}({\bm k}_1) \theta({\bm k}_2)\right\rangle\biggr]
\exp[i {\bm K} {\bm \cdot} {\bm R}] \,d {\bm  K}
\label{CC3}
\end{eqnarray}
are the third-order moment terms in ${\bm k}$ space appearing due
to the nonlinear terms, and
\begin{widetext}
\begin{eqnarray}
J_{j}({\bm k},{\bm R}) &=&  - i \int \Big[(\gamma-1)\, ({\bm k}_i + K_{i}/2 - Q_{i}) +Q_{i} \Big]
\, \left\langle u_i ({\bm k} + {\bm  K} / 2 - {\bm  Q})
u_j(-{\bm k} + {\bm  K}  / 2)\right\rangle \, \overline{T}({\bm Q})
\, \exp(i {\bm K} {\bm \cdot} {\bm R})\,d {\bm  K} \,d {\bm  Q} ,
\nonumber\\
\label{R14}\\
S({\bm k},{\bm R}) &=&  - i \int \biggl\{ \Big[(\gamma-1)\, ({\bm k}_j + K_{j}/2 - Q_{j}) + Q_{j} \Big]
\, \left\langle \theta(-{\bm k} + {\bm  K}  / 2) \, u_j({\bm k} + {\bm  K} / 2 - {\bm  Q}) \right\rangle
\nonumber\\
&& + \Big[(\gamma-1)\, (-{\bm k}_j + K_{j}/2 - Q_{j}) + Q_{j} \Big]
\, \left\langle \theta({\bm k} + {\bm  K}  / 2) \, u_j(-{\bm k} + {\bm  K} / 2 - {\bm  Q}) \right\rangle \biggr\}
\, \overline{T}({\bm Q}) \, \exp(i {\bm K} {\bm \cdot} {\bm R})\,d {\bm  K} \,d {\bm  Q} .
\nonumber\\
\label{D14}
\end{eqnarray}
\end{widetext}
\noindent
To derive Eq.~(\ref{C2}), we perform calculations in Eq.~(\ref{R14}) which are similar to those in Eqs.~(\ref{R4})--(\ref{R6}).
To determine $\langle \theta(\tilde {\bm k}_{1}) \, u_j(\tilde {\bm k}_{2})\rangle$ in Eq.~(\ref{D14}),
we use new variables:
\begin{eqnarray}
\tilde {\bm k} &=& (\tilde {\bm k}_{1} - \tilde {\bm k}_{2}) / 2 =
-{\bm k} + {\bm  Q} / 2 ,
\label{D16}\\
\tilde {\bm K} &=& \tilde {\bm k}_{1}
+ \tilde {\bm k}_{2} = {\bm  K} - {\bm  Q} ,
\label{D17}
\end{eqnarray}
where
\begin{eqnarray}
\tilde {\bm k}_{1} &=& -{\bm k} + {\bm  K} / 2  \;, \quad
\tilde {\bm k}_{2} = {\bm k} + {\bm  K} / 2 - {\bm  Q} .
\label{D15}
\end{eqnarray}
Since $|{\bm Q}| \ll |{\bm k}|$ and $|{\bm K}| \ll |{\bm k}|$, we use the Taylor expansion
\begin{eqnarray}
&& \left\langle \theta(\tilde {\bm k}_{1}) \, u_j(\tilde {\bm k}_{2}) \right\rangle = F_{j}(\tilde {\bm k}, \tilde {\bm K}) = F_{j}(-{\bm k},\tilde {\bm K}) + \frac{Q_m}{2}
{\partial F_{j}\over\partial \tilde k_m}
\nonumber\\
&& \quad + O({\bm Q}^2) = \left(1 - \frac{Q_m}{2}
{\partial \over\partial k_m}\right) F_{j}(-{\bm k},\tilde {\bm K}) + O({\bm Q}^2) .
\nonumber\\
\label{D18}
\end{eqnarray}
Similarly,
\begin{eqnarray}
&& \left\langle \theta(\tilde {\bm k}_{3}) \, u_j(\tilde {\bm k}_{4}) \right\rangle = \left(1 + \frac{Q_m}{2}
{\partial \over\partial k_m}\right) F_{j}({\bm k},\tilde {\bm K}) + O({\bm Q}^2) ,
\nonumber\\
\label{D19}
\end{eqnarray}
where
\begin{eqnarray}
\tilde {\bm k}_{3} &=& {\bm k} + {\bm  K} / 2  \;, \quad
\tilde {\bm k}_{4} = -{\bm k} + {\bm  K} / 2 - {\bm  Q} .
\label{D20}
\end{eqnarray}
Substituting Eqs.~(\ref{D18}) and~(\ref{D19}) into Eq.~(\ref{D14}),
neglecting the terms $O({\bm Q}^2; {\bm K}^2)$, and returning to the
physical space in the large-scale variables,
we obtain Eqs.~(\ref{C2}) and~(\ref{D21}).

To determine the turbulent heat flux and the level
of temperature fluctuations, we use the following identities for integration over $k$ in Eqs.~(\ref{C5}) and~(\ref{D22}):
\begin{eqnarray*}
\int_{k_0}^{k_\nu} \tau(k) \, \left[E(k) + \sigma_c \, E_c(k)\right] \, dk = \tau_0 \, (1 + \sigma_c) ,
\end{eqnarray*}

\begin{eqnarray*}
\int_{k_0}^{k_\nu} \tau(k)\, E(k) \, dk = \tau_0 \, \left[1 - {\tilde C_\sigma \, \sigma_c \over 2(1 + \sigma_c)} \right],
\end{eqnarray*}

\begin{eqnarray*}
\int_{k_0}^{k_\nu} \tau(k) \, E_c(k) \, dk = \tau_0 \, \left[1 + {\tilde C_\sigma \over 2(1 + \sigma_c)} \right],
\end{eqnarray*}

\begin{eqnarray*}
\int_{k_0}^{k_\nu} \tau(k) \, k^2 \, E_c(k) \, dk &=& {6 \tau_0 \over \ell_0^2} \left(1 + \sigma_c\right)^{-1} \biggl[{\rm Re}^{1/4}
\nonumber\\
&&+ {\sigma_c \over 4} \ln {\rm Re}\biggr] ,
\end{eqnarray*}

\begin{eqnarray*}
\int_{k_0}^{k_\nu} {d\tau(k) \over dk} \, E_c(k) \, k \, dk
&=& - {\tau_0 \, (q_c-1) \sigma_c \over 1 + \sigma_c}
\nonumber\\
&&\times \left[1 + {2(q-1) \over \sigma_c(q + q_c -2)} \right] ,
\end{eqnarray*}

\begin{eqnarray*}
\int_{k_0}^{k_\nu} \tau^2(k) \, k^2 \, E_c(k) \, dk = 4 f_c \, \left({\tau_0 \over \ell_0}\right)^2 \,  \left({\sigma_c \over 1 + \sigma_c} \right)^2 ,
\end{eqnarray*}

\begin{eqnarray*}
\int_{k_0}^{k_\nu} \tau^2(k) \, \left[E(k) + \sigma_c \, E_c(k)\right] \, dk = {4 \over 3} \, \tau_0^2 \, (1 + \sigma_c) ,
\end{eqnarray*}

\begin{eqnarray*}
\int_{k_0}^{k_\nu} \tau^2(k) \, E_c(k) \, dk = {4 \over 3} \, \tau_0^2 \, f_\ast \, \left({\sigma_c \over 1
+ \sigma_c} \right)^2 ,
\end{eqnarray*}

\begin{eqnarray*}
\int_{k_0}^{k_\nu} \tau^2(k) \, E(k) \, dk = {4 \over 3} \, \tau_0^2 \, (1 + \sigma_c) \, \left[1 - f_\ast \,  \left({\sigma_c \over 1 + \sigma_c} \right)^3 \right],
\end{eqnarray*}
where
\begin{eqnarray*}
f_\ast &=& 1 + {6 (q_c -1) \over \sigma_c (q + 2 q_c -3)} + {3 (q_c -1) \over \sigma_c^2
(2 q + q_c -3)} .
\end{eqnarray*}

\end{document}